# Cluster Formation and Percolation in Ethanol-Water mixtures

## Orsolya Gereben[1], László Pusztai[1,2,a]

[1]*Institute for Solid State Physics and Optics, Wigner Research Centre for Physics, Hungarian Academy of Sciences (Wigner RCP HAS), Konkoly Thege út 29-33, H-1121 Budapest, Hungary*

[2]*International Research Organization for Advanced Science and Technology (IROAST), Kumamoto University, 2-39-1, Kurokami, Chuo-ku, Kumamoto 860-855, Japan*

Results of systematic molecular dynamics studies of ethanol-water mixtures, over the entire concentration range, were reported previously to agree with experimental X-ray diffraction data. These simulated systems are analyzed in this work to examine cluster formation and percolation, using four different hydrogen bond definitions. Percolation analyses revealed that each mixture (even the one containing 80 mol % ethanol) is above the 3D percolation threshold, with fractal dimensions, $d_f$, between 2.6 to 2.9, depending on concentration. Monotype water cluster formation was also studied in the mixtures: 3D water percolation can be found in systems with less than 40 mol % ethanol, with fractal dimensions between 2.53 and 2.84. These observations can be put in parallel with experimental data on some thermodynamic quantities, such as the excess partial molar enthalpy and entropy.

---

[a] Corresponding author: L. Pusztai, Wigner RCP, HAS, H-1525 Budapest, Konkoly Thege út 29-33, Hungary; Tel/fax: + 36 1 392 2589; e-mail: pusztai.laszlo@wigner.mta.hu

# 1   INTRODUCTION

The structure of alcohol-water solutions has been studied widely by both experimental [1,2,3,4,5] and theoretical [6,7,8] methods. The emerging picture is still somewhat confusing and contradictory. Detailed discussions concerning these preliminaries have been provided in the introductory parts of our very recent publications [9,10,11], so these are not repeated here. In these recent works, H-bond connectivities [9], ring formation and statistics [10], and the 'lacunarity' [11] have been analyzed. To complete our extensive investigations on water-ethanol mixtures, here characteristics of cluster formation and percolation observable in these systems are presented. To the best of our knowledge, such properties have not been considered before for aqueous ethanol solutions.

Similarly to the aforementioned recent studies of water-ethanol mixtures [9,10,11], the basis of the present analyses is our preceding extensive molecular dynamics (MD) investigation [12]. There, a series of MD simulations for ethanol-water mixtures with 20 to 80 mol % ethanol contents, for pure ethanol and water was performed with one ethanol and three different water force fields. The primary aim of that work was to find the potential models that provide the best agreement with experimental X-ray diffraction data. In each mixture the OPLS-AA [13] potential was used for ethanol, in combination with three different water force fields, the rigid SPC/E [14] and TIP4P-2005 [15], as well the rigid polarizable SWM4-DP [16] ones. No single water force field could be identified that would provide the best agreement with experimental data over the entire concentration range: for higher ethanol contents the SWM-DP, whereas at lower ethanol concentrations the TIP4P-2005 potential provided the closest match.

In this work, detailed analyses of the hydrogen-bonded network are provided for the most successful simulations at each concentration, in terms of hydrogen bonded assemblies (clusters) of molecules. The number and size of the clusters strongly depend on the number of hydrogen bonds (HB) which, in turn, is determined by some (somewhat arbitrary) definitions of hydrogen bonds. As discussed (for instance) in our previous paper [9], there is not an exact rule for the definition, and several different approaches can be found in the literature. We have opted for purely geometric definitions: the O---H and O---O distances should fall into a certain distance range, and sometimes there is an additional constraint on the O-H…O angle, as well. A collection of such definitions are introduced and analyzed in Ref. [9]. Concerning earlier studies on ethanol-water mixtures, Noskov [17] applied an O---H cutoff of 2.4 Å and O-H…O angle>150º for defining H-bonds. Our criteria are comparable to these choices (see Ref. [9]). In our MD simulation investigation [12] the LOOSE condition for hydrogen bonding was based solely on distance ranges: the upper limiting values were defined by the first minima of the O−O, and by the second (i.e., first intermolecular) minima of the O−H PRDFs. On average, the "H-bonding" ranges were set to be between 2.4 and 3.6 Å for O−O and 1.4−2.7 Å for O−H pairs (see the Supporting Information Tables 1 and 2 in the preceding work for details [12]). A STRICT condition with the distance ranges of the LOOSE condition and with an additional angular restriction of O-H…O>120º was also applied; the 120º value was based on the definition of Chen et al. [18] and on our other previous work [19].

In the present paper, similarly to the immediate predecessor [9], four different H-bond definitions were applied, in order to explore how cluster formation might depend on the choice of the cutoff conditions. Apart from the LOOSE (L), and STRICT120 (S120) conditions (STRICT120 being the same as STRICT in the previous paper [12]), STRICT140 (S140) with O-H…O>140º and STRICT150 (S150) with O-H…O>150º are also applied here for each simulation result.

Percolation theory is frequently and successfully applied for explaining some of the unusual macroscopic features of hydrogen-bonded systems like water [20,21,22,23], methanol [24] and aqueous solutions [25,26]. Liquid water can be described by bond-percolation theory as a gel-like network with bent (and to some extent, broken) hydrogen bonds, well above the percolation threshold [20]. Dividing oxygen atoms into categories based on the number of their intact hydrogen bonds can be considered as a polychromatic correlated site percolation problem. The infinite hydrogen-bonded network in simulated water was found to contain patches of four-bounded water molecules with structures less ramified than that could be expected from their random distribution. This way, the anomalous behavior of, for example, the isothermal compressibility, constant-pressure and constant-volume specific heat and thermal expansion could be predicted and to some extent, explained [21]. The local density in the vicinity of these spatially correlated four-bonded patches was found to be lower than the global density [22].

According to random site percolation theory, infinite open clusters are true fractals at the percolation threshold with fractal dimensions $f_d$=2.53 for the 3D, and $f_d \cong$1.896 for the 2D case [27]. Based on the fractal dimension, a different percolation behavior was detected in tetrahydrofuran-water mixtures by Oleinikova et al. [25]. The predicted immiscibility gap was in good agreement with the concentration interval where both water and the solute were above their respective 3D percolation threshold. This suggested that totally miscible solutions can only exist if not both of the two compounds percolate in 3D [25]. Here, we wished to establish the percolation behavior of ethanol-water mixtures, so that later similar analyses may be conducted.

This paper is organized as follows: Section 2 briefly describes the computational methods used, while in Section 3, results on the cluster distributions and percolation properties are given in detail. Finally, Section 4 summarizes our findings.

## 2   METHODS

### 2.1   Molecular dynamics simulations

Details of the MD simulations were provided in our previous paper [12] and therefore here only a brief description is appropriate. All MD simulations were performed by the GROMACS 4.0 simulation package [28], in the NVT ensemble at $T$=293 K. The simulation length was 2000 ps in each case. Particle configurations in the production phase were collected 20 ps apart and in the end, 76 configurations were used for calculating average cluster sizes, connectivities and morphologies.

Calculations reported here are identified by their ethanol content and by the first letter of the water force field applied (see Table 1); for example, 'Et60S' refers to the mixture containing 60 mol % ethanol where the MD simulation was performed by using the SWM4-DP water force field (Table 1).

Table 1. Identifiers and densities for the MD simulations.

| $c_E$ (%)[a] | Simulation name | $n_E$[b] | $n_W$[c] | $\rho_E$ (Å$^{-3}$)[d] | $\rho_W$ (Å$^{-3}$)[e] | $\rho$ (Å$^{-3}$)[f] | $\rho_{liq}$ (g/cm$^3$)[g] |
|---|---|---|---|---|---|---|---|
| 100 | Et100 | 1331 | 0 | 0.0103 | 0.0000 | 0.0930 | 0.790 |
| 80 | Et80S | 1226 | 316 | 0.0096 | 0.0025 | 0.0940 | 0.799 |
| 60 | Et60S, Et60T | 1093 | 715 | 0.0088 | 0.0057 | 0.0960 | 0.816 |
| 40 | Et40T | 889 | 1328 | 0.0073 | 0.0109 | 0.0985 | 0.837 |
| 20 | Et20T | 571 | 2280 | 0.0047 | 0.0188 | 0.0990 | 0.841 |
| 0 | Et0S, Et0T | 0 | 3993 | 0.0000 | 0.0330 | 0.0990 | 0.987 |

[a] ethanol concentration in mol %.
[b] number of ethanol molecules.
[c] numer of water molecules.
[d] molecular number densities of ethanol molecules.
[e] molecular number densities of water molecules.
[f] atomic number density of all the atoms in the simulation cell.
[g] density of the liquids.

## 2.2 Cluster analyses

Analyses of the hydrogen-bonded clusters/network was performed by our own C++ computer code, specifically developed for this purpose, using a depth-first search algorithm for identifying clusters (see also Refs. [9,10,11]). The term 'cluster' is applied only for molecules connected by hydrogen bonds (i.e., lone molecules are not considered as clusters). Sometimes the distinction is made between cluster (no percolation) and network (percolation is present), but generally the term 'cluster' is used to describe assemblies of hydrogen bonded molecules, regardless of their percolation properties. During HB determination the simulation box was treated as an isolated system, as in the work of Geiger et al. [20], namely that only one instant of the periodic box – and consequently, each hydrogen bond – was used, even though hydrogen bonds were determined by using the minimum image convention.

Cycle perception was performed by the same software; detailed results can be found in Ref. [10].

## 2.3 Percolation

Percolation properties of each system have also been analyzed by the computer code mentioned above. For each cluster of each configuration of every system, it was determined whether the cluster in question was infinite; if it was then in how many dimensions. In general, a cluster is said to percolate if there is an infinite open cluster in at least one dimension. In this work the term 'network' is applied to a 3D percolating infinite cluster. Fractal dimensions have been calculated using the box-counting method, by the same software.

## 3   RESULTS AND DISCUSSION

Having located the H-bonds, clusters (molecules connected by HB-s) were identified; some of their statistical descriptors, as a function of the actual H-bond definition, have already been discussed in detail in Ref. [9].

### 3.1 Number of clusters

The average number of clusters for all the systems were calculated, and are shown in Fig. 1. Results for cluster sizes and distributions for the less strict HB conditions are shown, as well, in order to see the extent the choice affects the results.

For the same simulation with different HB conditions, the number of clusters increases with the increasing angular cutoff, as expected, similarly to the number of lone molecules. Regarding these values for the same HB condition for different concentrations, the number of clusters generally increases monotonically with increasing ethanol concentration (although there is a maximum around the Et40-Et60 region, especially for the stricter HB conditions). It is notable that the number of clusters for pure ethanol is substantially larger than it is for the mixtures, indicating a very different cluster formation, as discussed later.

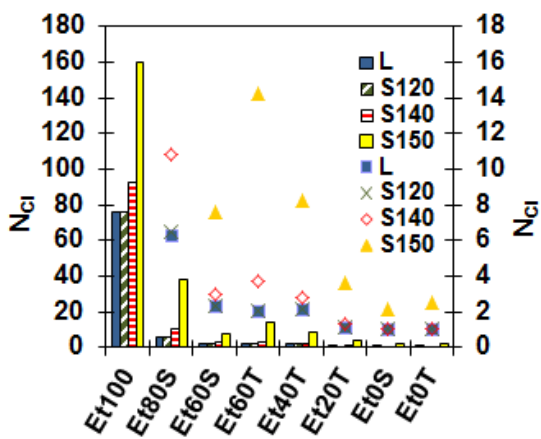

Fig. 1. Average numbers of clusters/configuration for ethanol-water mixtures, for pure ethanol and for pure SWM4-DP (denoted as 'Et0S') and TIP4P-2005 ('Et0T') water. Values belonging to the columns can be read on the left $y$-axis, while the smaller cluster numbers (particularly for water contents above 40 molar %) are represented by markers and can be read on the right (blown-up) $y$-axis, too, for the sake of clarity.

## 3.2 Cluster size distributions, percolation

The number of clusters versus the total number of molecules in the clusters is shown in Fig. 2(a). The 'small cluster size' range is enlarged in Fig. 2(b), and panels (c) and (d) depict the participation of water and ethanol molecules separately. (The values for the L and S120 HB conditions are almost identical, so only case L is shown.)

Distributions for pure ethanol for the various HB conditions are significantly different from the distributions of the water-containing systems. While for all the systems there are some small size clusters, the maximum size of these small clusters is the largest (~200 molecules) for pure ethanol, and decreases to 1 to 5, depending on the water force field and the HB condition when (even the smallest amount investigated here, 20 molar %, of) water is added.

There is one large agglomerate, with a characteristic cluster size, for all the water-containing systems (see Table 2). The majority of the molecules (86% for S150, 98% for the L condition for Et80S; 96 to ~100% for the other models and conditions) are in this main cluster (note that there is one main cluster in each particle configuration of the water-containing systems). The width of the maximum is characterized by the standard deviation (calculated from the sets of collected particle configurations; also given in Table 2); it is not changing much for the L, S120 and S140 HB conditions (although it is increased somewhat for the S150 HB condition; at the same time, the characteristic cluster size decreases somewhat).

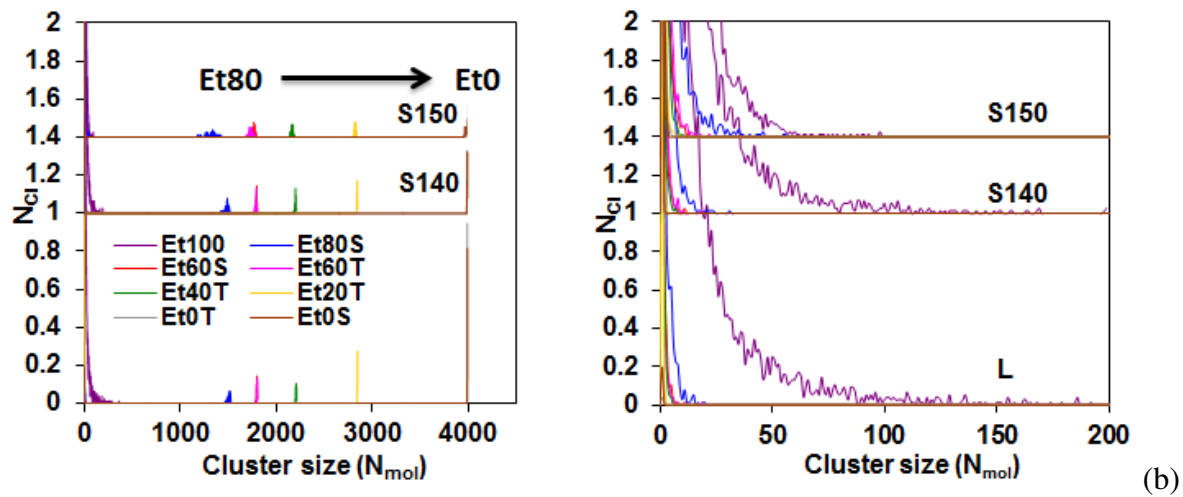

(a)

(b)

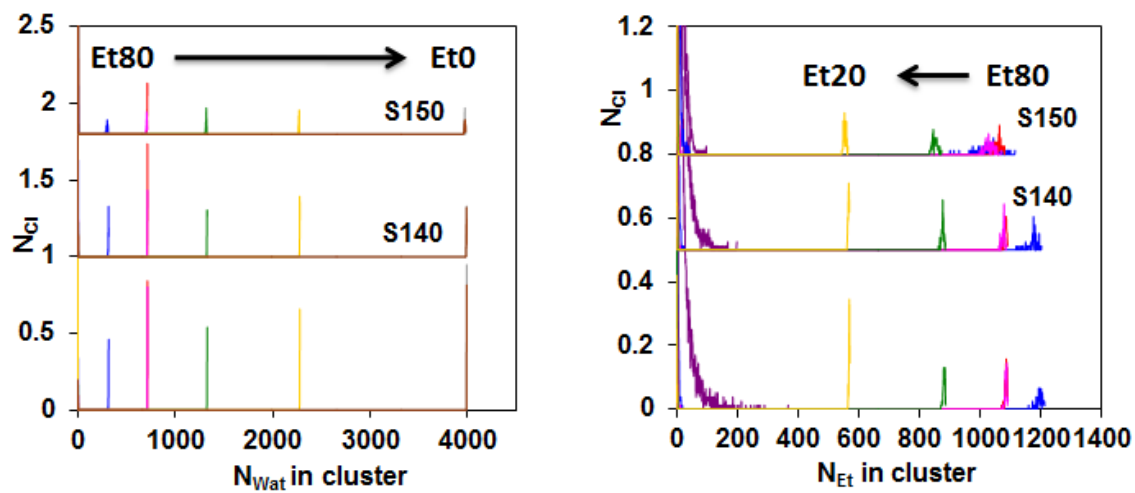

(c)

(d)

Fig. 2 (a) Cluster size (the total number of molecules in the cluster) distributions for all the simulations; (b) the same as (a), with an enlarged small cluster size region; (c) the number of clusters versus the number of water molecules in the clusters; (d) the number of clusters versus the number of ethanol molecules in the cluster. The legend is only displayed in panel (a); it is the same for the other panels.

Table 2. Characteristic cluster sizes (with standard deviations) for the percolating main clusters in the water-containing simulation boxes. (As there is not a percolating cluster in pure ethanol, that system is omitted.) Values in brackets are percentages of molecules that participate in the main cluster, as related to the total number of molecules or to the total number of ethanol molecules, respectively.

| Name | $N_{tot}$[a] | | | | $N_{Et}$[b] | | | |
|------|------|------|------|------|------|------|------|------|
| | L | S120 | S140 | S150 | L | S120 | S140 | S150 |
| Et80S | 1513±11 | 1512±12 | 1487±19 | 1329±46 | 1198±10 | 1197±11 | 1174±17 | 1030±40 |
| | (98.1) | (98.1) | (96.5) | (86.2) | (97.7) | (97.7) | (95.6) | (84.0) |
| Et60S | 1801±4 | 1801±4 | 1798±6 | 1775±11 | 1087±4 | 1087±4 | 1083±6 | 1062±10 |
| | (99.6) | (99.6) | (99.4) | (98.2) | (99.4) | (99.4) | (99.1) | (97.1) |
| Et60T | 1801±4 | 1801±4 | 1792±7 | 1738±16 | 1087±4 | 1086±3 | 1078±6 | 1031±14 |
| | (99.6) | (99.6) | (99.1) | (96.1) | (99.4) | (99.4) | (98.6) | (94.4) |
| Et40T | 2209±4 | 2209±4 | 2204±5 | 2169±10 | 882±3 | 882±3 | 877±4 | 851±8 |
| | (99.6) | (99.6) | (99.4) | (97.9) | (99.2) | (99.2) | (98.7) | (96.8) |
| Et20T | 2849±2 | 2849±2 | 2846±2 | 2825±6 | 569±1 | 569±1 | 567±2 | 555±4 |
| | (99.9) | (99.9) | (99.8) | (99.1) | (99.7) | (99.7) | (99.3) | (97.2) |
| Et0S | 3993±0.4 | 3993±0.5 | 3991±1 | 3974±5 | | | | |
| | (~100) | (~100) | (99.95) | (99.5) | | | | |
| Et0T | 3993±0.3 | 3993±0.3 | 3992±1 | 3974±4 | | | | |
| | (~100) | (~100) | (99.95) | (99.5) | | | | |

[a]total number of molecules.

[b]number of ethanol molecules.

Percolation analysis was performed and revealed the presence of an infinite open cluster, with 3D percolation, in all the water-containing systems. Due to the percolation in the water-containing systems the actual characteristic cluster size would depend on the size of the simulation box, but the percentages of the molecules involved (given in brackets in Table 2) and the trend how it is changing with concentration and HB condition would remain similar. On the other hand, percolation analysis in Et100 (pure ethanol) revealed that only in cases of the (less strict) L, S120 and S140 HB conditions can one single 1D infinite cluster be found, in one configuration of the configuration ensemble. This equals to an occurrence probability of 1.3%, way below the percolation threshold; note that no percolation at all could be detected if the strictest (S150) HB condition was used.

Fractal dimensions for the percolating clusters were also calculated and are displayed in Table 3. The values increase from $d_f$=2.60 to 2.99 with increasing water concentration, and all of them are above $d_f$=2.53, which is the 3D percolation threshold predicted by random site percolation theory [27]. If the fractal dimensions are plotted against the ethanol concentration (given in molar %), a nearly linear behavior can be found; only the value for the Et80S simulation with the S150 HB condition falls somewhat lower, see Fig. 3.

Table 3: Average fractal dimensions with standard deviations for the percolating main clusters of the ethanol-water mixtures and pure water.

| Name | L | S120 | S140 | S150 |
|------|------|------|------|------|
| Et80S | 2.67±0.006 | 2.67±0.006 | 2.66±0.008 | 2.60±0.018 |
| Et60S | 2.73±0.004 | 2.73±0.004 | 2.73±0.004 | 2.72±0.004 |
| Et60T | 2.74±0.003 | 2.74±0.003 | 2.73±0.004 | 2.72±0.006 |
| Et40T | 2.81±0.003 | 2.81±0.003 | 2.81±0.003 | 2.80±0.004 |
| Et20T | 2.90±0.002 | 2.90±0.002 | 2.90±0.002 | 2.90±0.003 |
| Et0S | 2.99±0.001 | 2.99±0.001 | 2.99±0.001 | 2.99±0.001 |

Et0T    2.99±0.001    2.99±0.001    2.99±0.001    2.99±0.001

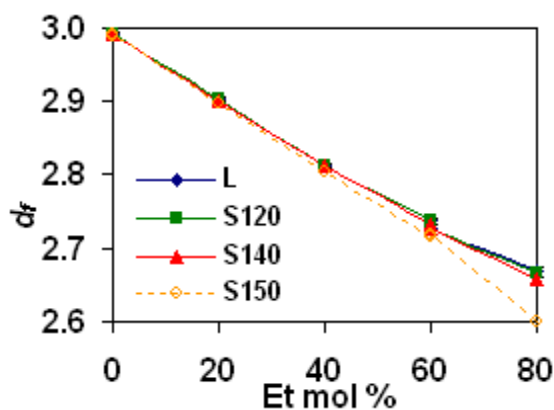

Fig. 3. Fractal dimension versus the ethanol concentration (in molar %) for the percolating systems.

Considering the number of water and ethanol molecules in the clusters (Fig. 2 (c) and (d) and Table 2) the following observations can be made:

1)    There is not much variation in terms of the number of water molecules in the clusters across the individual particle configurations, whereas the number of ethanol molecules shows almost the same standard deviation as the total number of molecules. That is, water molecules are more stably 'glued' to the main cluster than ethanol molecules.

2)    The participation of ethanol molecules in the main cluster is between 84.0 and 99.7% (if we omit the lowest value, which occurs in the Et80S system using the S150 condition, they change between 94.4 and 99.7%), while for the water molecules it is 94.8 to ~100%. Again, water molecules appear to like being in the main cluster to a somewhat larger extent than ethanol moleules.

The occurrence of purely ethanol and purely water clusters among the small isolated clusters was also determined; results are shown in Fig. 4, as the ratio (in %) of the number of isolated clusters. While for ethanol the ratio increases from ~7% to ~76% with increasing ethanol concentration, it declines quickly for pure water clusters from 14% to 0% with decreasing water concentration. The ratio of molecules in pure clusters, as compared to the total number of molecules of their own type, is small: the largest value is ~7% and quickly decreases to 0.02% for ethanol with decreasing ethanol concentration.

The maximum number of molecules in a purely ethanol cluster is 13 to 15 for the Et80S systems, but only 2 to 5 for the most diluted Et20T system. The ratio of water molecules involved in pure water clusters is <0.05%, and most of such clusters consist of two (and at most of three) molecules. This shows that ethanol has a larger affinity to form small, pure clusters, especially in the high ethanol concentration range, than water. At the same time, at low concentration (~20% ethanol) there is no evidence of substantial chainlike cluster formation, in contradiction with the suggestions of Sato et al. [5].

Based on the cluster analysis we can conclude that the aggregation properties of pure ethanol are different from the rest of the systems: there are many, smaller sized clusters, and no percolation can be found there. Water-containing mixtures have aggregation characteristics similar to water, as even the Et80S system is above the 3D percolation threshold. These mixtures contain a binary infinite open cluster (network), and some smaller clusters can be found, as well. It can therefore be established that the percolation threshold for ethanol-water mixtures lies somewhere in the 80 to 100 mol % ethanol region.

It is worth pointing out that in terms of the excess activation enthalpy and entropy of these mixtures, a change of the slope of both of these quantities versus the ethanol concentration was observed at ~90 mol % ethanol concentration by Sato et al. [5]. The excess partial molar enthalpy and entropy for water also change their slope and increase in this composition region. As these quantities are considered to be the consequences of multiple interactions between ethanol and water, it is tempting to suggest that such changes can be related to the percolation threshold of binary clusters in the system.

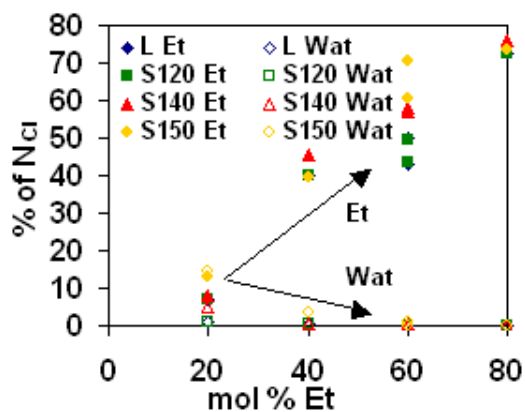

Fig. 4. The ratio of pure ethanol and water clusters among the small clusters.

### 3.3 Monotype cluster formation

So far all types of hydrogen bonds were considered during cluster formation; from this point on, we will only consider hydrogen bonds between the same types of molecules, as this can reveal information about the possible internal structure of the clusters. Also, such analyses may be able to provide information on micro-phase separation. It has to be noted that these monotype clusters are mostly parts of larger, binary clusters, and only a portion of them are pure isolated clusters; statistics concerning the pure isolated clusters were given in the previous section.

The cluster/network analysis was performed similarly as for the binary clusters, but first only for the ethanol molecules, and then only for the water molecules in the particle configurations. The analysis was conducted for all the HB conditions, and gave similar results concerning observable trends. Here results only for the strictest (and possibly, the most reasonable) S150 HB condition will be discussed.

Statistics concerning the monotype clusters are shown in Fig. 5, where the respective concentrations of the investigated molecule types are given; that is, values for monotype ethanol clusters of the 20 mol % ethanol containing simulation are compared to values of monotype water cluster formation found in the simulated system with 80 mol % ethanol (20 mol % water), and so on.

The number of molecules involved in monotype hydrogen bonding is roughly the same for the 20 mol % ethanol and 20 mol % water simulations, although the total number of ethanol molecules in the latter is almost twice as large. This points toward a much higher ratio of involvement (~65%) for water than ethanol (~25% , see Fig. 5(a)). At higher molar fractions the values for water increase more rapidly than for ethanol: they show a behavior closer to linear for ethanol, while reaching nearly 100% hydrogen bonding already at 60 mol % for water. Average cluster sizes (Fig. 5 (b) upper panel) increase near linearly from ~2 to ~8.

Considering the number of clusters (see Fig. 5(b) lower panel), both ethanol and water have similar values at 20 mol % (~60); ethanol presents a maximum curve, having the largest number of monotype clusters at the concentration of 80 %. On the other hand, the number of water clusters decreases monotonically, except for the value of 40 mol % water when the TIP4P-2005 force field is used. There is a large difference between values of the TIP4P-2005 and SWM4-DP force field simulations for the 40 mol % water concentration: the SWM4-DP force field brings about higher monotype hydrogen bonding capability, and therefore higher average cluster size and lower number of clusters, in accordance with our previous findings based on the atomic connectivities. Therefore we can conclude that water has higher affinity to form monotype clusters (to have at least one water neighbor), which can be attributed only partially to the larger number of water molecules in the systems.

Already the (small) number of (water) clusters suggested that for ≥60% water content, there might be water percolation in the system. This could be confirmed by performing percolation analysis for the monotype water clusters at each concentration: indeed, three-dimensional percolation was found for ≥60% water concentrations (characteristic cluster sizes are shown in

Table 4). In case of system Et40T with the S150 HB condition, 95% of the configurations had 3D, while 5% 2D percolation; this is over the 3D percolation threshold. (The less strict HB conditions had 100% 3D percolation.) The monotype percolating characteristic cluster size is ~45-55% of the respective binary characteristic cluster size shown earlier in Table 2 for the Et40T and ~80% for the Et20T simulations.

Fractal dimensions were calculated for the monotype percolating clusters, as well; they are quoted in Table 5. The values are between $d_f$=2.53 and 2.6 for the Et40T, and 2.83 and 2.84 for the Et20T simulation, depending on the HB condition. $d_f$=2.53 was found for the Et40T S150 system with 95% 3D percolation, not much above the percolation threshold.

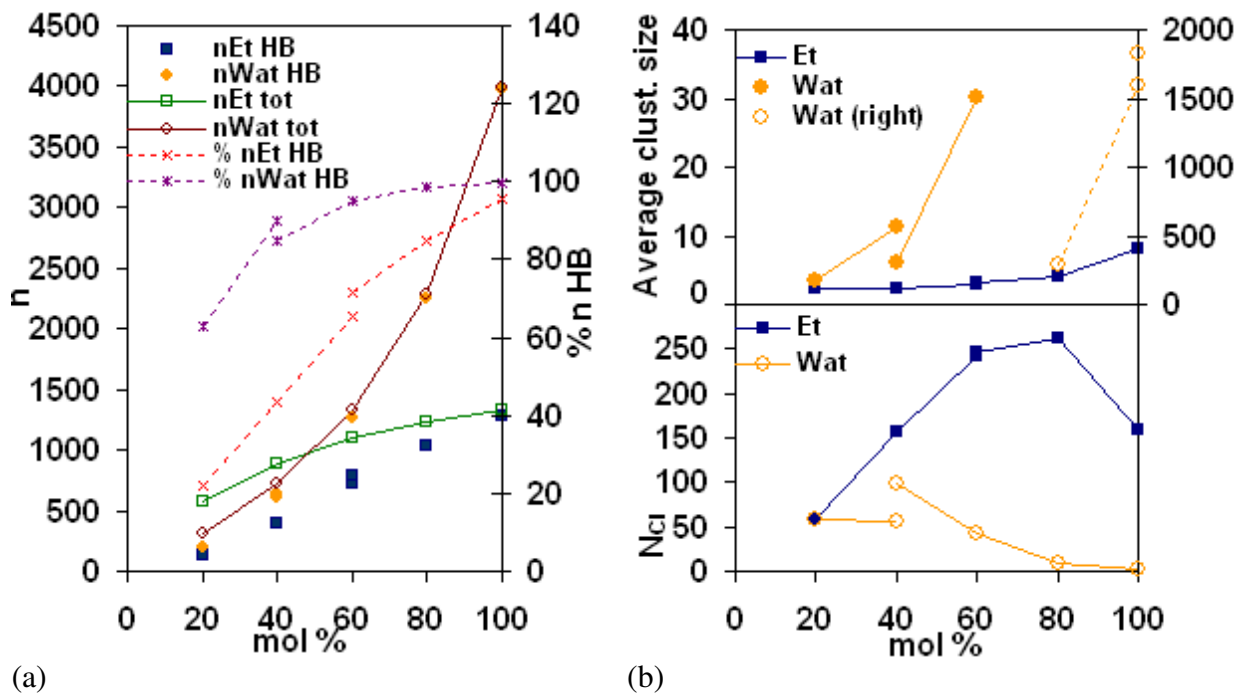

(a)                                                                (b)

Fig. 5. Statistics concerning the monotype cluster formation for simulations with different concentrations and using the S150 HB conditions. (a) Total number of ethanol (nEt tot) and water (nWat tot) (left *y*-axis), number of ethanol (nEt HB) and number of water molecules (nWat HB) involved in monotype hydrogen bonding (left *y*-axis), and the percentage of ethanol (% nEt HB) and water (% nWat HB) atoms involved in monotype hydrogen bonding (right *y*-axis) plotted against the respective molar fractions. b) Upper panel: average monotype cluster sizes and lower panel: number of monotype clusters ($N_{Cl}$) plotted against the respective molar fractions. The 80 and 100% values for the water clusters are plotted against the right *y*-axis in the upper panel, and represented by empty markers for easier distinction. Only data points created by the same force field are connected by a line.

Table 4. Characteristic sizes of the percolating monotype water clusters; standard deviations are also shown. Values in brackets are the percentages of the molecules contained in the monotype percolating cluster compared to the total number of water molecules. (There are no percolating ethanol clusters, as discussed in the previous section.)

| Name | L | S120 | S140 | S150 |
|---|---|---|---|---|
| Et40T | 1223±19 | 1223±20 | 1200±28 | 988±57 |
| | (92.1) | (92.1) | (90.4) | (74.4) |
| Et20T | 2271±4 | 2271±4 | 2266±5 | 2235±9 |
| | (99.6) | (99.6) | ((99.4) | (98.0) |
| Et0S | 3993±0.4 | 3993±0.5 | 3991±1 | 3974±5 |
| | (~100) | (~100) | (99.95) | (99.5) |
| Et0T | 3993±0.3 | 3993±0.3 | 3992±1 | 3974±4 |
| | (~100) | (~100) | (99.95) | (99.5) |

Table 5. Average fractal dimensions for the monotype 3D percolating clusters (both Et60 simulations are mentioned).

| Name | L | S120 | S140 | S150 |
|---|---|---|---|---|
| Et60S | 1.73±0.50 | 1.73±0.50 | 1.73±0.49 | 1.71±0.36 |
| Et60T | 1.60±0.20 | 1.60±0.20 | 1.58±0.27 | 1.54±0.25 |
| Et40T | 2.60±0.01 | 2.60±0.01 | 2.59±0.01 | 2.53±0.07 |
| Et20T | 2.84±0.001 | 2.84±0.001 | 2.84±0.001 | 2.83±0.001 |
| Et0S | 2.99±0.001 | 2.99±0.001 | 2.99±0.001 | 2.99±0.001 |
| Et0T | 2.99±0.001 | 2.99±0.001 | 2.99±0.001 | 2.99±0.001 |

From the behavior of the excess partial molar enthalpy and entropy in the 10 to 18 mol % ethanol concentration range, calculated from dielectric relaxation measurement data, Sato et al. [5] suggested that the hydrogen bond network of water is disrupted by the ethanol molecules, and 'percolation nature of water will no longer be present in the region >0.1 Et mol fraction'. Our molecular dynamics results clearly oppose this suggestion, as even in the case of 40 mol % ethanol content 3D water percolation was found, regardless of the water force field and the HB condition.

Sato et al. [5] found changes in the slope of the excess activation free energy, enthalpy and entropy around 50 mol % ethanol concentration. They suggested that some structural changes occur in this region. Nishikawa et al. [1] suggested that concentration fluctuations reach their maximum at ~40 mol % ethanol, based on small angle X-ray scattering experiments. Ben-Naim [6] found that at ~47 mol % ethanol concentration the average affinity between water molecules reaches its maximum, while between ethanol and water molecules, reaches its minimum. All of these findings can be connected to the fact that, according to our simulation data, the percolation threshold for water can be found somewhere between 40 and 60 mol % ethanol concentration in ethanol-water mixtures. Furthermore, Oleinikova et al. [25] reported the 2D percolation threshold for water in tetrahydrofuran-water mixtures at 53 mol % water concentration, and the 3D threshold at 63 mol % water, which are not far from our results.

## 4   CONCLUSIONS

(1) Pure ethanol has different cluster forming properties from the rest of the systems: many smaller sized clusters can be found up to 100 to 370 molecules, depending on the HB condition applied. In this liquid, no percolation occurs.

(2) On the other hand, all the studied ethanol-water mixtures have a 3D percolating infinite main cluster, and therefore they form a network. The size of the percolating main cluster is increasing with increasing water concentration. Smaller clusters, <50 molecules, can be found beside the main cluster in these systems; their size and number is decreasing with increasing water concentration. For pure water (almost) all the molecules are in the main cluster.

(3) Ethanol is more likely to be found in smaller binary or pure clusters, or to remain as solitary molecules (without hydrogen bonding).

(4) Fractal dimensions of the infinite percolating clusters are increasing from $d_f$=2.6 to 2.9 for the ethanol-water mixtures and $d_f$=2.99 for water, regardless of the choice of the water force field or HB condition.

(5) If only the monotype cluster formation is considered (these clusters are mostly part of larger binary clusters), then 3D water percolation is found in the Et20T and Et40T systems, putting the percolation threshold for water in ethanol-water mixtures between the 40 and 60% ethanol content. This is significantly higher than the ethanol content of 10%, suggested before by Sato et al. [5]. There are conjectures based on various experiments [1,5,6] that some structural changes occur around ~50% ethanol content: this is consistent with our findings. Fractal dimensions for the infinite percolating water clusters are between 2.53 and 2.84; these values are above the 3D percolation threshold.

(6) Regarding the choice of the hydrogen bond condition, there are obviously some quantitative differences in terms of the cluster number and size distributions and fractal dimensions, and the exact value of the percolation threshold would be affected, too. However, the statement that 3D percolation exists in these systems in the investigated concentration range is still valid. Therefore qualitatively our findings do not depend on the choice of the applied hydrogen bond condition; it seems to be prudent to use the strictest S150 HB condition, as in this case the connectivities appear to be the most realistic.

(7) Monotype cluster formation (and also, lacunarity analysis [11]) indicates that although microscopic phase separation might occur in these systems, macroscopic phase separation is not observable (as it is well known). Although 3D water percolation is present in the ≤40 mol % ethanol concentration region, no percolation can be found in pure ethanol and therefore, 3D ethanol percolation cannot occur in the mixtures, either. This may explain why ethanol-water mixtures remain miscible over the entire concentration region, as immiscibility seems to occur if both of the components are percolating in 3D separately (cf. Ref. [25]).

**ACKNOWLEDGMENTS**

This work was supported by the National Research, Development and Innovation Agency (NRDIO; in Hungarian: NKFIH, Hungary), under contract No. SNN 116198.